\documentclass[prl,aps,showpacs,twocolumn]{revtex4}
\usepackage{amsmath, amsthm,times}
\usepackage{epsfig}
\usepackage{amssymb}
\usepackage{txfonts}
\usepackage{graphicx}
\usepackage{bm}

\DeclareMathSymbol{\C}{\mathbin}{AMSb}{"43}
\DeclareSymbolFont{AMSb}{U}{msb}{m}{n}

\begin{document}

\sloppy

\title{Generating Entangled Photons from the Vacuum by Accelerated Measurements:\\
Quantum Information Theory Meets the Unruh-Davies Effect}

\author{Muxin Han,\ \ S. Jay Olson,\ \ Jonathan P. Dowling}

\affiliation{Hearne Institute for Theoretical Physics, Department of Physics and Astronomy, Louisiana State
University, Baton Rouge, LA 70803, USA}


\begin{abstract}

Building on the well-known Unruh-Davies effect, we examine the effects of projective measurements and quantum communications between accelerated and stationary observers.  We find that the projective measurement by a uniformly accelerated observer can excite real particles from the vacuum in the inertial frame, even if no additional particles are created by the measurement process in the accelerating frame. Furthermore, we show that the particles created by this accelerating measurement can be highly entangled in the inertial frame, and it is also possible to use this process to generate even maximally entangled two-qubit states by a certain arrangement of measurements. As a byproduct of our analysis, we also show that a single qubit of information can be perfectly transmitted from the accelerating observer to the inertial one. In principle, such an effect could be exploited in designing an entangled state generator for quantum communication.

\end{abstract}

\maketitle

It has become well-known in the thirty years since the discovery of the Unruh-Davies effect that the concept of particle is dependent on an observer's state of motion \cite{unruh}, somewhat analogous to the way in which distance and time become dependent on an observer's state of motion, with the introduction of special relativity.  In the Unruh-Davies effect, unitarily evolving particle detectors will respond to the same inertial quantum vacuum state in very different ways, depending on the acceleration of the detectors.  Following this surprising result, a great deal has been written regarding its implications for quantum field theory in particular and physics in general \cite{unruhwald,Kay,wald1}, with the debates sometimes centering on what is meant by the reality of the Unruh particles, or invoking different starting assumptions and arguments that lead to similar results \cite{dowling,Ford,Lin}.

The Unruh-Davies effect itself is a statement about the unitary transformation of the quantum vacuum between independent observers, and does not take into account the effects of communication between stationary and accelerated observers, or the effects of non-unitary projective measurements occurring between the reference frames. These might seem to be important concepts for comparing the different experiences of the different quantum observers.  Recently, however, some interesting ideas have begun to emerge in the field of relativistic quantum information and quantum entanglement studying well-known staples such as quantum teleportation in the context of accelerating observers \cite{Alsing1,reznik,vanEnk,bradler}.

Here we continue to develop this approach, and analyze the effects of projective measurement on Unruh particles in an accelerating frame, combined with the communication of the result (via a purely quantum communication channel) to an inertial observer.  Remarkably, we find that such projective measurements in the accelerating frame can create real particles in the inertial frame \---- \emph{even if no additional particles are created by the measurement process in the accelerating frame.} By this, we explicitly mean the following: if we have two independent observers in the vacuum, one at rest and the other in uniform acceleration, the projective measurements made by the accelerating observer will create particles detectable by the inertial observer \---- she effectively gains access to real particles via measurements on what was initially nothing. Furthermore, if some certain projective measurements are chosen by the accelerated observer (e.g. measuring the particle number), we have found that the inertial frame particles generated by accelerated measurement are always highly entangled, representing a generator for entanglement resources \---- all available simply by measuring the vacuum.

\begin{figure}
\begin{center}
\includegraphics[height=6cm, width=6cm, angle=0]{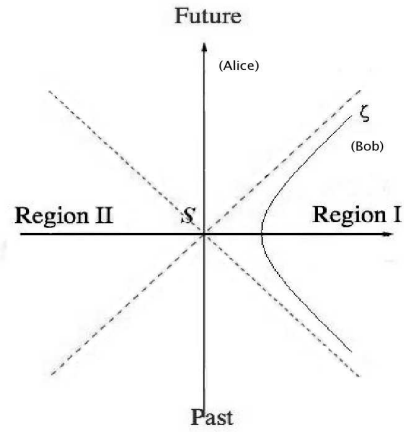}
\caption{Alice and Bob in Minkowski spacetime.}
\end{center}
\end{figure}

In the following analysis, we consider the linear-optical case and neglect the photon polarization, so that we are effectively concentrating our discussion on a massless scalar field. However, all of our discussions and results can be straightforwardly applied to the the case of photons with polarization. Before jumping into the details of the calculation, we first outline the physical process involved:

The processes we propose can be set up by two observers together with their associated detectors. One observer (Alice) is inertial while the other one (Bob) is uniformly accelerating with acceleration $a$. The Fock spaces associated with each of them are denoted by $\mathcal{F}_{A}$ and $\mathcal{F}_{B}$ respectively. The initial state is the vacuum state of standard quantum field theory $|0\rangle_A\in\mathcal{F}_{A}$, where Alice sits. Then Bob makes a standard von-Neumann projective measurement \footnote{Practically, the standard von-Newmann measurement can be realized by the so called, quantum nondemolition (QND) measurement \cite{dowling2}.} on the photon number of any single mode $\omega_1$ and gets an activation of his detector via the Unruh-Davies effect. From the collapse of quantum state, Alice will see photon creation in her frame and, we will show that the photons created in the inertial frame are always entangled between certain different modes. 

We analyze the effect in 3+1-dimensional Minkowski spacetime. First of all, the world-line of Bob in the right wedge (region I) is generated by the time-like vector field
\begin{eqnarray}
\zeta^a=a\Bigg[X(\frac{\partial}{\partial
T})^a+T(\frac{\partial}{\partial X})^a\Bigg]\nonumber
\end{eqnarray}
where $a$ is the acceleration and $(T, X)$ are global inertial coordinates. Hence we can construct a quantum field theory in region I using $\zeta^a$ as the time translation vector, and also in region II by using $-\zeta^a$ since $\zeta^a$ is past-directed in region II. Then we have the well-known Bogoliubov transformation \cite{Davies,bao,wald1}:
\begin{eqnarray}
Va^I_{\omega}V^{-1}&=&(1-e^{-\frac{2\pi\omega}{a}})^{1/2}[b_{\omega}+e^{-\frac{\pi\omega}{a}}b'^\dagger_{\omega}],\nonumber\\
Va^{II}_{\omega}V^{-1}&=&(1-e^{-\frac{2\pi\omega}{a}})^{1/2}[b'_{\omega}+e^{-\frac{\pi\omega}{a}}b^\dagger_{\omega}],\label{bogoliubov}
\end{eqnarray}
where the operator $V:\ \mathcal{F}_{B}\rightarrow\mathcal{F}_{A}$ is the S-matrix connecting the bosonic Fock space of Bob to the bosonic Fock space of Alice. Then $a^I_{\omega}$ ($a^{II}_{\omega}$) operating on $\mathcal{F}_B$ is the annihilation operator associated with the solution $\psi^I_{\omega}$ ($\psi^{II}_{\omega}$) to the Klein-Gordon equation, which vanishes in region II (region I) and oscillates harmonically with frequency $\omega>0$ with respect to the accelerating time translation \footnote{Rigorous speaking, $\psi^I_{\omega}$ and $\psi^{II}_{\omega}$ are a basis of normalized wavepackets of positive frequency solutions, whose frequencies are peaked sharply about $\omega$ (see e.g. Refs \cite{Hawking,wald2}).}. The annihilation operators $b_{\omega}$ and $b'_{\omega}$ on $\mathcal{F}_A$ are associated with the one-photon states
\begin{eqnarray}
\Psi_\omega&=&(1-e^{-\frac{2\pi\omega}{a}})^{1/2}[\psi^I_{\omega}+e^{-\frac{\pi\omega}{a}}{\psi^{II*}_{\omega}}],
\nonumber\\
\mathrm{and} \ \ \ \ \
\Psi'_\omega&=&(1-e^{-\frac{2\pi\omega}{a}})^{1/2}[\psi^{II}_{\omega}+e^{-\frac{\pi\omega}{a}}{\psi^{I*}_{\omega}}],\nonumber
\end{eqnarray}
respectively.

Following the derivation of Unruh-Davies effect, we can express the standard Minkowski vacuum state $|0\rangle_A\in\mathcal{F}_A$ as a quantum state in $\mathcal{F}_B$, which can be detected by Bob \cite{wald,wald1}
\begin{eqnarray}
V^{-1}|0\rangle_A&=&c_0\prod_\omega\Bigg[\sum_{n_\omega=0}^{\infty}\frac{1}{n_\omega!}\Big(e^{-\frac{\pi\omega}{a}}
a^{I\dagger}_{\omega}a^{II\dagger}_{\omega}\Big)^{n_\omega}\Bigg]\ |0\rangle_B\label{unruhstate}
\end{eqnarray}
where $c_0$ is the overall normalization constant and $|0\rangle_B$ is the vacuum state in the accelerating frame.

Suppose that Bob in region I measures the photon number on an arbitrary frequency $\omega_1$ and obtains the number $m$. Then the state will be projected to the component which has the photon number distribution $n_{\omega_1}=m$ on the frequency $\omega_1$, since the experimental results in region I and region II are correlated in the sense of Eq.(\ref{unruhstate}). The resultant state after his measurement is denoted by $|\Phi\rangle_B$ with the expression:
\begin{eqnarray}
|\Phi\rangle_B&=&c_1\frac{1}{m!}\Big[e^{-\frac{\pi\omega_1}{a}}
a^{I\dagger}_{\omega_1}a^{II\dagger}_{\omega_1}\Big]^{m}\nonumber\\
&&\prod_{\omega\neq\omega_1}\Bigg[\sum_{n_\omega=0}^{\infty}\frac{1}{n_\omega!}\Big(e^{-\frac{\pi\omega}{a}}
a^{I\dagger}_{\omega}a^{II\dagger}_{\omega}\Big)^{n_\omega}\Bigg]\ |0\rangle_B,\label{Phi'}
\end{eqnarray}
where we have renormalized the state by the factor $c_1$.

Then we switch back to the inertial frame to figure out what Alice obtains. Recall that the S-matrix $V$ makes the connection between the Fock spaces of Bob and Alice, so after Bob finishes his work, Alice should have the state $V|\Phi\rangle_B$. Since the terms in Eq.(\ref{Phi'}), with respect to the frequency $\omega_1$ and other frequencies are factorized, one can see from Eq.(\ref{unruhstate}) that all the components with respect to the frequencies $\omega\neq\omega_1$ will be transformed back to be the vacuum state. After some calculations for the Bogoliubov transformation, we obtain
\begin{eqnarray}
V|\Phi\rangle_B
&=&c_2(b^\dagger_{\omega_1}+e^{-\frac{\pi\omega_1}{a}}b'_{\omega_1})^m(b'^\dagger_{\omega_1}+e^{-\frac{\pi\omega_1}{a}}b_{\omega_1})^m\nonumber\\
&&\sum_{n=0}^\infty\frac{1}{n!}(-e^{-\frac{\pi\omega_1}{a}}b_{\omega_1}^\dagger
b'^\dagger_{\omega_1})^n\ |0\rangle_A\nonumber\\
&=&c_3\sum^\infty_{l=0}\sum^m_{q=0}(-1)^lK^m_{ql}(\omega_1)\ |q+l;q+l\rangle_A,\label{result1}
\end{eqnarray}
where
\begin{eqnarray}
K^m_{ql}(\omega)&=&\delta_{q,0}e^{-\frac{l\pi\omega}{a}}\ \ \ \ \ \ \ \ \ \ \ \ \ \ \ \ \ \mathrm{for}\ \ m=0\nonumber\\
K^m_{ql}(\omega)&=&\frac{m!}{q!(m-q)!}\prod_{i=1}^m(l+i)e^{-\frac{m-q+l}{a}\pi\omega}\ \ \ \ \mathrm{otherwise},\nonumber
\end{eqnarray}
and the number state $|\ i;j\ \rangle_A$ means that there are $i$ and $j$ inertial frame photons in the modes $\Psi_{\omega_1}$ and $\Psi'_{\omega_1}$ respectively. Therefore, we can see that Alice obtains real photons, since Bob's projective measurements have modified the original state. A simple way to view this effect can be given as follows. In the standard Unruh-Davies effect the inertial vacuum is unitarily transformed into a thermal bath Eq.(\ref{unruhstate}) for the accelerating observer. Since any unitary transformation is uniquely invertible \---- \emph{only} a thermal bath in the accelerated frame transforms back to an inertial vacuum. Once the accelerating observer makes a measurement, he destroys the purely thermal nature of the field. Hence, it should not be surprising that when the collapsed thermal field is transformed back, that the inertial observer no longer sees vacuum. Furthermore, one can check the non-separability of the state Eq.(\ref{result1}) via the Positive Partial Transpose (PPT) criterion \cite{HHH1,peres}. The partial transposed density matrix $\rho^{PT}$ is obtained from $\rho:=V|\Phi\rangle_{BB}\langle\Phi|V^\dagger$ by $_A\langle i;j\ |\ \rho^{PT}|\ k;l\rangle_A=_A\langle k;j\ |\ \rho\ |\ i;l\rangle_A$, so it can be expressed as:
\begin{eqnarray}
&&\rho^{PT}\nonumber\\
&=&\sum_{l,l'=0}^\infty\sum_{q,q'=0}^m(-1)^{l+l'}K_{ql}^mK_{q'l'}^m|q'+l';q+l\rangle_{AA}\langle q+l;q'+l'|\nonumber
\end{eqnarray}
up to a normalization constant. It can be checked straightforwardly by definition that the operator $\rho^{PT}$ is not non-negative, i.e. $\langle\psi|\ \rho^{PT}|\psi\rangle$ fails to be non-negative for all $|\psi\rangle$ in its Fock space, so $\rho^{PT}$ must have a negative eigenvalue. Thus by the PPT criterion, the photon state Eq.(\ref{result1}) is non-separable between the two modes $\Psi_{\omega_1}$ and $\Psi'_{\omega_1}$. Therefore Bell's inequality should be violated since it is a pure state. So the inertial frame photons generated by the accelerating measurement is in an entangled state. The entanglement is between the photon numbers in the modes $\Psi_{\omega_1}$ and $\Psi'_{\omega_1}$.

Moreover, if we suitable arrange the projective measurements and properly design a quantum communication protocol, it is even possible for us to obtain an almost maximally entangled two-qubit state (EPR state) in the inertial frame, which is suitable for use in quantum cryptography \cite{ekert}. First let's recall the accelerating projective measurement made by Bob. However, instead of measuring a single frequency as before, we let Bob measure the photon numbers for two different frequencies $\omega_1$ and $\omega_2$ and obtain $m_1$ and $m_2$ respectively. Then the corresponding projected state $|\Phi\rangle_B$ changes to
\begin{eqnarray}
|\Phi'\rangle_B&=&c'_1\frac{1}{m_1!}\Big[e^{-\frac{\pi\omega_1}{a}}
a^{I\dagger}_{\omega_1}a^{II\dagger}_{\omega_1}\Big]^{m_1}\frac{1}{m_2!}\Big[e^{-\frac{\pi\omega_2}{a}}
a^{I\dagger}_{\omega_2}a^{II\dagger}_{\omega_2}\Big]^{m_2}\nonumber\\
&&\prod_{\omega\neq\omega_1,\omega_2}\Bigg[\sum_{n_\omega=0}^{\infty}\frac{1}{n_\omega!}\Big(e^{-\frac{\pi\omega}{a}}
a^{I\dagger}_{\omega}a^{II\dagger}_{\omega}\Big)^{n_\omega}\Bigg]\ |0\rangle_B.\nonumber
\end{eqnarray}
When we switch back to Alice's frame, it is also clear that Alice's vacuum is excited, and the photons in the inertial frame are created by Bob's accelerating measurement. The expression for the inertial frame photon state is obtained via the Bogoliubov transformation in the same way as we did before:
\begin{eqnarray}
V|\Phi'\rangle_B
&=&c'_2(b^\dagger_{\omega_1}+e^{-\frac{\pi\omega_1}{a}}b'_{\omega_1})^{m_1}(b'^\dagger_{\omega_1}+e^{-\frac{\pi\omega_1}{a}}b_{\omega_1})^{m_1}\nonumber\\
&&(b^\dagger_{\omega_2}+e^{-\frac{\pi\omega_2}{a}}b'_{\omega_2})^{m_2}(b'^\dagger_{\omega_2}+e^{-\frac{\pi\omega_2}{a}}b_{\omega_2})^{m_2}\nonumber\\
&&\prod_{i=1,2}\sum_{n_i=0}^\infty\frac{1}{n_i!}(-e^{-\frac{\pi\omega_i}{a}}b_{\omega_i}^\dagger
b'^\dagger_{\omega_i})^{n_i}\ |0\rangle_A\label{result2}
\end{eqnarray}
Since the terms with respect to $\omega_1$ and $\omega_2$ are completely factorized, Eq.(\ref{result2}) essentially is a tensor product of two versions of Eq.(\ref{result1}) with different modes. (Obviously, if Bob measures photon numbers of $n$ frequencies, Alice will obtain a $n$-fold tensor product of Eq.(\ref{result1}).) So our previous observation of entanglement is also applied to the photon state, Eq.(\ref{result2}), in which the entanglement not only takes place between the modes $\Psi_{\omega_1}$ and $\Psi'_{\omega_1}$, but also between $\Psi_{\omega_2}$ and $\Psi'_{\omega_2}$. On the other hand, one can rewrite the expression of Eq.(\ref{result2}) to be a linear combination of the photon number basis:
\begin{eqnarray}
V|\Phi'\rangle_B
&=&c'_3\sum^\infty_{l,l'=0}\sum^{m_1}_{q=0}\sum^{m_2}_{q'=0}(-1)^{l+l'}K^{m_1}_{ql}(\omega_1)K^{m_2}_{q'l'}(\omega_2)\nonumber\\ 
&&|q+l,q'+l';q+l,q'+l'\rangle_A,\label{result3}
\end{eqnarray}
where the number state $|i,j;k,l\rangle_A$ means that there are $i$ photons in the mode $\Psi_{\omega_1}$, $j$ photons in the mode $\Psi_{\omega_2}$, $k$ photons in the mode $\Psi'_{\omega_1}$, and $l$ photons in the mode $\Psi'_{\omega_2}$.

Given the initial photon state generated by the accelerating measurement, one may perform some selections on it to obtain some more interesting quantum states. For example, we let Alice make one more projective measurement on the total photon number. If the result of the measurement is two, Eq.(\ref{result3}) is projected to its two photon component, which is a two-qubit entangled state:
\begin{eqnarray}
|\Theta\rangle_A&=&[K^{m_1}_{10}-K^{m_1}_{01}](\omega_1)K^{m_2}_{00}(\omega_2)|1,0;1,0\rangle_A\nonumber\\
&+&[K^{m_2}_{10}-K^{m_2}_{01}](\omega_2)K^{m_1}_{00}(\omega_1)|0,1;0,1\rangle_A,\label{output}
\end{eqnarray}
which are non-separable and entangled as is easily shown from the Positive Partial Transpose criterion. In addition, one may notice that the coefficients in the above state are different only up to a simple switch of labels between 1 and 2. Therefore, Eq.(\ref{output}) is almost a maximally entangled EPR state when the following two conditions are satisfied:  (i) Bob detects the same number $m=m_1=m_2$ for the two different frequencies $\omega_1$ and $\omega_2$ \footnote{The successful probability for this $\prod_{i=1,2}e^{-\frac{2m\pi\omega_i}{a}}(1-e^{-\frac{2\pi\omega_i}{a}})$ may be read from Eq.(\ref{unruhstate}).} and (ii) the difference between $\omega_1$ and $\omega_2$ is sufficiently small, or effectively, Bob's acceleration $a$ is sufficiently large. However, in the above processes of generating entangled two photonic qubits, Alice may need to clarify the degree of entanglement for the resultant state $|\Theta\rangle_A$, in order to properly use them. Then she should know from Bob his acceleration, the value of the frequencies, and how many photons in each frequency. So a channel of information flow from Bob to Alice is necessary in the application. Such a one-way information flow can be realized by the signal photons created from Bob in his frame on the background of thermal spectrum, Eq.(\ref{unruhstate}), because a qubit of information can always be perfectly transfered from Bob to Alice via a signal photon with frequency $\omega_0\neq\omega_1,\omega_2$, i.e. when Bob creates a signal photon in his frame, by the transformation: 
\begin{eqnarray}
Va^{I\dagger}_{\omega_0}V^{-1}|0\rangle_A=(1-e^{-\frac{2\pi\omega_0}{a}})^{1/2}b_{\omega_0}^\dagger|0\rangle_A,\nonumber
\end{eqnarray}
Alice always receives the signal without any degradation due to the thermal spectrum of photons in the accelerating frame \cite{Alsing1}. And the qubit of information carried by the signal photon is invariant under the Bogoliubov transformation between the two reference frames.

At last, we conclude our discussion by noting a few interesting implications of this effect: \\
(i) Since the particle interpretation depends on the observer's state of motion, the above effect suggests that the interpretation of a projective measurement on a particle should also depend on the observer's state of motion. Although it seems counter intuitive that particle creation should be a result of making a projective measurement of particle number, the analysis above is exactly a demonstration of such a phenomenon. The detector does not create any new particles in its own reference frame (it simply measures particles already present), but it does create particles in another reference frame, and the observers in the other frame are perfectly free to detect and use them in every real sense. Since the created inertial frame particles are always highly entangled, our scheme is a generator of entangled resources.\\
(ii) It should be noted that after the projective measurements made by the accelerating observer, we not only have physical photons in the inertial frame, but we also obtain a non-zero energy-momentum tensor of the quantum field, in contrast to the vanishing energy-momentum tensor of the original vacuum state. Thus one can see that the accelerated projective measurement is really a process of \emph{emitting} energy and momentum \---- and thus the act of measurement acquires an additional input of energy from the accelerating agent. \\
(iii) It is also remarkable that the proposed quantum-optical-signal communication can perfectly transmit a qubit of information from the accelerating observer to the inertial observer, simply by creating a single-photon qubit over the thermal background of Unruh particles, Eq.(\ref{unruhstate}). However, the reverse qubit transmission from the inertial observer to the accelerating observer is not a satisfactory means of communication, since the accelerating detector can be activated even when no qubits have been sent by the inertial observer \cite{Alsing1}.  This was not a problem for our analysis, of course, because we only consider a one-way flow of information from the accelerating frame to the inertial frame.

\section*{Acknowledgments}

This work is supported by the funding from the Army Research Office and the Disruptive Technologies Office. Muxin Han would also like to acknowledge the support of NSFC 10205002 and NSFC 10675019. All the authors would like to thank Hugo Cable, Pavel Lougovski, Barry Sanders, John Sipe, Sai Vinjanampathy, Christoph Wildfeuer, Ulvi Yurtsever, and Hongbao Zhang for useful discussions and suggestions.

\end{document}